Submitted to J. Phys. Chem. C, 6/30/2020# Potential Energy Landscape of CO Adsorbates on NaCl(100) and Implications in Isomerization of Vibrationally Excited CO

Jun Chen,[1] Seenivasan Hariharan,[2] Jörg Meyer,[2] and Hua Guo[1,*]

[1]Department of Chemistry and Chemical Biology, University of New Mexico, Albuquerque, New Mexico, 87131, USA

[2]Leiden Institute of Chemistry, Gorlaeus Laboratories, Leiden University, P.O. Box 9502, 2300 RA Leiden, The NetherlandsTOC graphic

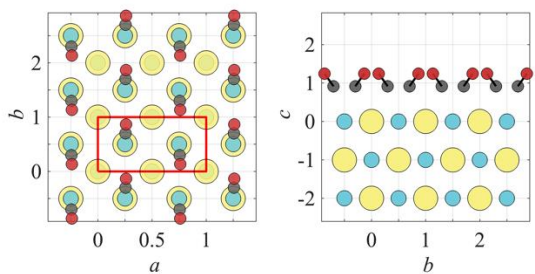

*: corresponding author, hguo@unm.edu1

**Abstract**

Several full-dimensional potential energy surfaces (PESs) are reported for vibrating CO adsorbates at two coverages on a rigid NaCl(100) surface based on first principles calculations. These PESs reveal a rather flat energy landscape for physisorption of vibrationless CO on NaCl(100), evidenced by various C-down adsorption patterns within a small energy range. Agreement with available experimental results is satisfactory, although quantitative differences exist. These PESs are used to explore isomerization pathways between the C-down and higher energy O-down configurations, which reveal a significant isomerization barrier. As CO vibration is excited, however, the energy order of the two isomer changes, which helps to explain the experimental observed flipping of vibrationally excited CO adsorbates.

**I.     Introduction**

CO adsorption on surfaces of salt crystals such as NaCl has provided an ideal proving ground to study vibrational energy flow among weakly interacting molecules with well-defined separation and orientation.[1] The interaction between the CO adsorbate and the ionic surface is dominated by weak electrostatic (ES) and van der Waals (vdW) interactions, as evidenced by a relatively low adsorption energy and a small CO frequency shift.[2, 3] For CO in its ground vibrational state, the most stable configuration features carbon down (C-down) adsorption at the $Na^+$ site,[4] thanks to the slightly negative charge in carbon. The adsorbate-substrate interaction is augmented by weak but long-range inter-adsorbate interactions. The energy landscape of the CO adsorption is thus quite flat near the equilibrium adsorption geometry.[5-9] At high temperatures, the CO monolayer (ML) on NaCl(100) has a two-dimensional 1×1 lattice with CO oriented perpendicular to the surface, while this structure transforms to a 2×1 lattice with titled CO at temperatures lower than 35 K.[10-12]

In a series of pioneering papers about thirty years ago, Chang and Ewing demonstrated that laser excitation of CO molecules adsorbed on cold NaCl(100) surfaces to their low-lying vibrational states ($v$=1) can lead to facile energy transfer among the adsorbates, resulting in some highly excited CO molecules, which can be detected from their spontaneous emission.[13-15] This so-called vibrational energy pooling has since attracted much attention from both theoretical[16-19] and experimental fronts.[20, 21] The underlying basis for the CO molecules to climb the vibrational ladder is its anharmonicity, which gives rise to a small exoergicity for processes such as CO($v$) + CO($v'$) → CO($v$-1) + CO($v'$+1).[22] This near-resonant energy transfer between two vibrationally excited CO adsorbates is very efficient, even when they are separated by a few lattice sites, and the small energy release can be soaked up by phonons of the cold surface.[20] This vibrational exchange process dominates over direct CO vibrational relaxation (CO($v$) → CO($v$-1)), which requires many phonons to accept the large energy release associated with the loss of a vibrational quantum, as the Debye frequency of the NaCl surface is merely 223 cm$^{-1}$. The reverse endoergic ladder-descending process is essentially dampened on cold surfaces by the Boltzmann factor, thus negligible at low temperatures. Interestingly, such vibrational energy exchange is also operative in the gas phase,[23-25] for applications such as supersonic CO laser.[26] The ladder climbing is realized in the gas environment by collisions, by the same principle as mentioned above.

Very recently, high resolution spectra of the CO on cold NaCl(100) system have been measured suggesting the energy pooling can result in population of CO vibrational states as high as $v$=40.[20, 21] More interestingly, these highly vibrationally excited CO are observed to flip from a C-down geometry to an O-down geometry, thus providing an interesting example of double-well systems coupled to a condensed medium.[21] To better understand the energy pooling and the resulting isomerization processes, one needs to map out the global potential energy surface (PES) for CO adsorption on NaCl with the C-O vibrational coordinate included, which is the main objective of this work. To this end, we report several global PESs for CO adsorption on a rigid NaCl(100) surface based on analytical representations of density functional theory (DFT) points using a machine learning method. Several different functionals were tested and two vdW-corrected ones were used in generating the DFT points. We emphasize that these PESs differ from previous empirical ones[6, 7, 16] in that the interaction is computed using first principles methods and the PESs assume no particular form. These PESs allows us to gain insights into the energy landscape for adsorption, vibration, and isomerization of CO molecules on the surface. This work is organized as follows. The next section (Sec. II) outlines the DFT methods and fitting of the PESs. The results are presented and discussed in Sec. III. A summary is given in Sec. IV.

## II. Methods

### A. Density Functional Theory

All planewave DFT calculations were performed with the VASP (Vienna *Ab initio* Simulation Package) code.[27, 28] Three different exchange-correlation functionals were used in the calculations. The first is the generalized gradient approximation (GGA) type functional of Perdew, Burke, and Ernzerhof (PBE).[29] To include dispersion interactions ignored in the PBE functional, two types of vdW-corrected DFT methods, the D3 method with Becke-Johnson damping (PBE-D3(BJ)),[30, 31] and the revised many-body dispersion energy method including fractionally ionic contributions to the polarizability (PBE-MBD@rsSCS/FI),[32, 33] were employed.

The NaCl(100) surface was modeled by a two-layer slab and the CO overlayer is simulated in $p(1\times1)$ (and $p(2\times1)$) or $p(2\times2)$ surface unit cells, corresponding to 1 ML or 1/4 ML coverage, respectively. The optimized NaCl lattice has a lattice constant of 5.697 Å with PBE, 5.583 Å with PBE-D3(BJ) and 5.664 Å with (PBE-MBD@rsSCS/FI). The experimental lattice constant of 5.640 Å[34] differs by less than 0.06 Å from these theoretical values. In order to systematically focus on the differences of the exchange-correlation functionals in this study, we have thus used the experimental value for the rigid NaCl slab in all our calculations. The vacuum space between the periodic slabs in the Z direction was set to 15 Å. The interaction between ionic cores and electrons was described by projector-augmented wave (PAW) potentials.[35] A Monkhorst-Pack $k$-points grid mesh[36] of $3\times3\times1$ was used, and the planewave expansion was truncated at a kinetic energy of 700 eV. The convergence properties of slab layers and $k$-points are provided in Figure S1 in Supporting Information (SI).

### B. Data Sampling and PES Fitting

Since the full-dimensional PES is designed to cover the entire configuration space, it requires points near the adsorption equilibrium as well as the desorption asymptote for vibrationally excited CO. The data sampling was started by extracting the geometries and energies from *ab initio* molecular dynamics (AIMD) trajectories of CO on the NaCl(100) surface. These trajectories were launched at 6.0 Å from the surface with CO($v$ = 0 - 40), directed toward the surface along the surface normal. The CO molecule is randomly oriented, but with zero rotational angular momentum. A geometric criterion of 0.1 Å based on the root mean square deviation (RMSD) of the Euclidean distance between two points was applied to exclude points that were too close to each other. A primitive PES was constructed based on the first batch of approximately 3000 geometries. Additional points were sampled by running quasi-classical trajectory (QCT) calculations with various initial conditions ($v_{CO}$ = 0 - 40) on this primitive PES. A new point was included into the data set if it satisfied the aforementioned geometric criterion and an additional energetic criterion given by RMSD of energies predicted by five different fits of the data. Specifically, the energetic criterion excludes those points with almost the same predicted results from different fits, as the fit can be considered to be converged at these locations. Then the PES was updated using the new data set. This procedure was repeated iteratively, and the PES was considered to be fully converged if no point below 2.5 eV can be found from a new batch of trajectories, with the energetic criterion set to 10 cm$^{-1}$. This systematic data sampling approach has been validated in various gas-phase and molecule-surface interaction systems.[37-39] Finally, 16196 symmetry unique points were sampled.

Feed forward neural networks (NNs) with two hidden layers were employed to fit the six-dimensional PESs. All the 16196 points were projected to the irreducible triangle of the NaCl(100) surface unit cell in advance, as shown in Figures 1(a) and 1(b). Any points outside this symmetry unique region can be obtained by the symmetry of the surface. To ensure the symmetry and continuity of the boundary, those points located near the boundary of the irreducible triangle were expanded using symmetry operations beyond the symmetric unique region near the boundary, as illustrated by the red points in Figures 1(a) and 1(b), which makes a total of 20611 points for NN fitting. A vector containing 6 fractional coordinates (for the $p(1\times1)$ surface unit cell) and a bond length of CO, *i.e.*, [$X_C$, $Y_C$, $Z_C$, $X_O$, $Y_O$, $Z_O$, $r_{CO}$], was used as the input layer of the NN functions. The inclusion of the redundant C-O distance helps to converge the results better. 50 neurons were used for each hidden layer, after testing different numbers of neurons. The structure of a NN function can thus be denoted as 7-50-50-1, which contains 3001 parameters. The NN functions were fitted with the expanded data set divided into two sets (90% for the training set and 10% for the validation set) using the Levenberg-Marquardt algorithm[40] with an early-stopping method.[41] The final PES was an average on five best fits to further reduce random errors.

## III. Results

### A. Adsorption Configurations and Energies

We investigate adsorption of CO on NaCl(100) with three different surface unit cells. The models with the $p(1\times1)$ and $p(2\times1)$ CO layer are both for the 1 ML coverage, while the $p(2\times2)$ model explores the behavior of CO at a lower (1/4 ML) coverage. The first two scenarios have been observed experimentally as the high- and low-temperature phases of the CO adlayer on NaCl(100).[10, 11] The choice of the $p(2\times2)$ model is designed to understand the CO energetics on NaCl(100) with minimal CO-CO interactions, as the distance between two adjacent CO in this coverage is 7.98 Å. This model is important to extract the interaction PES between an isolated CO and the NaCl(100) surface, as discussed below.

The local and global potential minima for the $p(1\times1)$ surface unit cell (1 ML) were optimized using different DFT methods, and the corresponding energies, geometries and CO vibrational frequencies are listed in Table 1. The coordinates used to describe these geometries are depicted in Figures 1(e) and 1(f). In addition to three tilted C-down minimum energy structures, a tilted O-down minimum has been located at a significantly higher energy. They are denoted as $s1$, $s2$, $s3$, $s4$ hereafter, and their images are shown in Figure 2. For the results using the PBE functional, the global minimum corresponds to the $s1$ geometry, which has a tilt angle of 30.19°, a very small azimuthal angle of 1.96°, and a significant lateral displacement of the carbon atom, $dX_C = 0.6588$ Å and $dY_C = 0.0447$ Å, from the top Na$^+$ site. This configuration features a CO molecule shifted from one surface Na$^+$ site to a neighboring Na$^+$ site, as illustrated in Figure 2. In contrast, the other two C-down minima have azimuthal angles nearly 45°, shifted from the Na$^+$ site to the neighboring Cl$^-$ site. The tilt angle of $s2$ is somewhat smaller than $s1$, while the $s3$ is very close to the perpendicular configuration. The O-down minimum $s4$ has a tilt angle of 143.39° (= 180° - 36.61°), and also large shifts in both $X$ and $Y$ directions, which resembles the C-down $s2$ configuration. In addition, there exists a perpendicular O-down minimum, denoted as $s5$, but only with the PBE functional. These adsorption configurations are qualitatively the same for the two vdW-corrected functionals, PBE-D3(BJ) and PBE-MBD@rsSCS/FI, although the C-down $s3$ configuration from PBE-D3(BJ) is more tilted and shifted from the Na$^+$ site.

We have also investigated the CO structure in $p(2\times1)$ unit cell, which is known to exist on NaCl(100) at temperatures below 35 K.[10,11] Like the $p(1\times1)$ model described above, this model corresponds to 1 ML coverage, but it allows two CO in the surface unit cell to orient differently. Optimization results show that the two CO in a unit cell have exactly the same values of $r_{CO}$, $dZ$ and tilt angles, and the only difference is the sign of the azimuthal angle. As shown in Table 2 and Figure 3, seven minimum energy structures, five C-down and two O-down, have been found. Among them, the C-down $s1$ and $s2$ structures are identical to those in the $p(1\times1)$ (1 ML) model. The relaxation of the relative orientation of the two CO molecules in the unit cell allows additional adsorption patterns, leading to two minimum energy structures that were first discussed in the work of Vogt and Vogt.[12] The antiparallel minimum C-down $s3$ is the global minimum in the $p(2\times1)$ model with all the three functionals, which can be generated from $s1$ with opposite azimuthal angles and lateral displacements $dY$ for the two CO molecules on the neighboring $Na^+$ sites. On the other hand, the herringbone minimum $s4$ has close absolute values of geometry parameters with $s2$ but different signs in the azimuthal angle and slightly different $dZ$. Another minimum, denoted as $s5$, which is similar to $s4$ but has smaller azimuthal angles, has also been located. Comparing with $s2$, $s4$ has a slightly higher energy and $s5$ is energetically more favorable. In addition, a herringbone O-down minimum $s7$ has been located in the $p(2\times1)$ model, which has an almost undistinguishable adsorption energy from that of the minimum $s6$. The energetical similarities between $s1/s3$ and between $s2/s4/s5$ in the $p(2\times1)$ model indicate that the $p(1\times1)$ model should be good enough in describing the energy landscape of CO adsorbate on NaCl(100), apart from the orientational differences.

The C-down minima found in the $p(2\times1)$ (1 ML) model were also reported by Boese and Saalfrank.[9] These authors also found other tilted configurations with irregular or spiral oriented CO (denoted as T/I or T/S, respectively) in larger unit cells.[9] For each size of the unit cell, the antiparallel configuration remains to be the most energetic favorable. Interestingly, the C-down minima obtained from the $p(1\times1)$ and $p(2\times1)$ models can be identified in the local structures of T/I or T/S configurations. The combination of these minima in Tables 1 and 2, which have similar energies, results in the richness and diversity of configurations in larger unit cells.

The geometry of the calculated global minimum (antiparallel, $s3$) can be compared with experimental geometric information obtained from LEED,[12] which is also listed in Table 2. Both the CO bond-length and the height ($dZ$) are in good agreement with experimental estimation. However, its tilt angle is larger than the experimental value. In fact, the experimental tilt and azimuthal angles are closer to the herringbone structure ($s4$), which has a slightly higher energy. As discussed below, the energy landscape is quite flat and these two structures ($s3$ and $s4$) might be sufficiently close in energy to interconvert, even at low temperatures. Thus, we conclude that the overall agreement with the experiment is satisfactory, albeit with some quantitative uncertainties. A likely source of the uncertainties is the inaccuracy of the DFT functional, which might be responsible for the noticeable structure differences in Table 2. More accurate electronic structure methods, such as those based on correlated wave functions,[42, 43] are need to provide a quantitative comparison for this floppy system. Another promising approach is to express the interactions in a pairwise form in which the interactions are obtained with a high-level *ab initio* method.[44] In addition, some of the uncertainties may also derive from the finite experimental temperature (25 K) , which might require a free-energy simulation to make the direct comparison with experiment, due to the floppy nature of the system. Furthermore, as already suggested by Vogt and Vogt,[12] it might be necessary to re-analyze the LEED data by including anisotropic thermal displacements which our calculations can provide. We are currently investigating whether this can improve the agreement with the experiment.

For the $p(2\times2)$ surface unit cell, corresponding to a lower CO coverage of 1/4 ML, the local and global minima have also been determined and their information is listed in Table 3. Their images are given in Figure 2. Interestingly, the tilt angles and lateral shifts from two vdW-corrected functionals are all smaller compared to the corresponding geometries from Tables 1 and 2, presumably due to the larger CO-CO distances and hence weaker CO-CO interactions. With the PBE functional, the global minimum, the C-down *s*3, has a very small tilt angle (0.6°) and almost no lateral shift, located at the top site of Na$^+$ site. In contrast, results obtained with both the PBE-D3(BJ) and PBE-MBD@rsSCS/FI functionals predict the C-down *s*1 configuration as the global minimum, similar to that of *p*(1×1) model. No tilted O-down minimum was found using the PBE-MBD@rsSCS/FI functional.

From Tables 1-3, it is clear that all C-down minima in all three scenarios have very similar energies, suggesting a relatively flat potential landscape. These are all typically physisorption, judging from the small shifts in bond-length and frequency from the free CO. Figure 4 plots the adsorption potential energy curves (PECs) as a function of distance from C atom of CO to the NaCl surface, calculated with different methods for the *p*(1×1) (1 ML) and *p*(2×2) (1/4 ML) models. The minima of these PECs correspond to the global minimum C-down configurations shown in Tables 1 and 3. The energy zero is defined as the sum of the energy of the clean surface and the energy of a single free CO molecule placed in a large box (20×20×20 Å). The adsorption energies in the *p*(1×1) (1 ML) model are generally larger than those in the *p*(2×2) (1/4 ML) model, due to the attractive CO-CO interactions. For the same reason, the potential energies in the *p*(1×1) (1 ML) model do not vanish when the CO molecule desorbs, except for the PEC from the PBE functional, which lacks vdW interactions between CO molecules. Comparing with the adsorption energy (-1504 cm$^{-1}$ or -18.0 kJ/mol) obtained from the experiments by Richardson *et al.*,[3] the PBE functional grossly underestimates due apparently to lack of the vdW interaction, but the PBE-D3(BJ) functional substantially overestimates. On the other hand, the PBE-MBD@rsSCS/FI functional is in good agreement with experiment[3] and the high-level QM:QM embedding result of -17.8 kJ/mol by Boese and Saalfrank,[9] presumably because of a balanced description of the vdW corrections. We note in passing that all of the functionals considered here yield zero-point energy corrections that destabilize the adsorption energies by about +2.4 kJ/mol for the $p(1\times1)$ and $p(2\times1)$ structures, which is in qualitative but not quantitative agreement the estimate of 4-5 kJ/mol by Richardson *et al.*.[3] Finally, the equilibrium adsorption height, which is defined as the distance between C and the surface, is measured to be 2.59 Å,[12] which is reproduced with reasonable accuracy by all three methods, as shown in the figure.

**B. Potential Energy Surfaces**

Because of the poor performance of the PBE functional in reproducing the adsorption energy, it is not considered in future discussions. Instead, both vdW-corrected functionals, PBE-D3(BJ) and PBE-MBD@rsSCS/FI, are used for the construction of the six-dimensional PESs, describing the adsorption of one CO molecule in the $p(1\times1)$ and $p(2\times2)$ surface unit cells, respectively. The twelve-dimensional PES for $p(2\times1)$ was not attempted, because the potential landscape is qualitatively similar to that of $p(1\times1)$, as discussed above. The 16196 configurations used for PES fitting are distributed evenly on the irreducible zone of the NaCl(100) surface, as shown in Figure 1(a) for the $p(1\times1)$ model and Figure 1(b) for the $p(2\times2)$ model. Distribution of the data points in the height of the CO molecule to the surface, the bond length of CO as well as the total potential energy are illustrated in Figure 1(c) and Figure 1(d). Four PESs have been constructed, denoted hereafter as (a) 2×2-MBD, (b) 2×2-D3, (c) 1×1-MBD and (d) 1×1-D3, with the fitting RMSE values of 3.68, 3.60, 5.71 and 5.68 meV, respectively. The distribution of fitting errors is displayed in Figure S2 as a function of potential energy, in which the four PESs perform similarly. The PESs can be obtained from the corresponding author upon request.

Figure 5 shows the contour plots of the PESs as a function of the tilt angle and height of CO (defined as $dZ = (dZ_C+dZ_O)/2$, as illustrated in Figure 1(f)) to the NaCl(100) surface. The potential energies were calculated with the CO bond fixed at its equilibrium length (1.132 Å), and with the other three dimensions (azimuthal angle, $dX$ and $dY$) optimized. Figures 5(a) and 5(b) display the 2×2-MBD and 2×2-D3 PESs between the slightly tilted C-down minimum ($s1$) and the O-down minimum ($s5$ for 2×2-MBD and $s4$ for 2×2-D3), denoted in the figure as "A" and "B". In Figures 5(c) and 5(d) the 1×1-MBD and 1×1-D3 PESs are shown between two tilted C-down ($s1$) and O-down ($s4$) minima, denoted in the figure as "C" and "D". The minimum energy structures and energies are consistent with those listed in Tables 1 and 3. In both coverages, the O-down geometry has a significantly higher energy than the C-down geometry, and the conversion between the two has to surpass a significant barrier.

As shown in Tables 1 and 3, the PESs are relatively flat with several C-down minima within a small energy range. Figure 6 shows the PES of the CO adsorbate on different adsorption sites, with the C-O bond fixed at its equilibrium and other three coordinates ($dZ$, the tilt and azimuthal angles) optimized. Due to the similarity between PBE-MBD@rsSCS/FI and PBE-D3(BJ) functionals, only the results on 2×2-MBD and 1×1-MBD PESs are displayed. The C-down $s1$/$s2$/$s6$ structures on the 2×2-MBD PES, and $s1$/$s2$/$s3$ structures on the 1×1-MBD PES can be clearly observed. For the 2×2-MBD PES, the $s1$/$s2$ and $s6$ configurations have similar energies, with differences smaller than 5 cm$^{-1}$ between each other. As a result, the PES is very flat on top and around the Na$^+$ site. In contrast, on the 1×1-MBD PES, the C-down $s1$/$s2$ configurations have potential energies almost 80 cm$^{-1}$ lower than $s3$, thus the adsorbed CO molecules are generally shifted from the Na$^+$ site.

To shed light on the energy landscape for a vibrationally excited CO adsorbate on NaCl(100), Figure 7 shows the PESs between a stretched CO molecule with the NaCl(100) surface, with the bond length $r_{CO}$ fixed at 1.596 Å and the other three dimensions optimized. The $r_{CO}$ corresponds to the outer turning point of the CO vibrational state $v = 20$. There are two minima, denoted as F/E and G/H in the figure. The lower minima as marked by "F" and "H", have tilt angle near 90°, with the $O^{\delta-}$ atom adsorbed to the surface $Na^+$ atom and the $C^{\delta+}$ atom adsorbed to the surface $Cl^-$ atom, as illustrated in Figure S3. The higher minima, as marked by "E" and "G" in Figures 7(a) and 7(c), have also adsorption geometry nearly parallel to the surface plane with the $O^{\delta-}$ atom adsorbed to the surface $Na^+$ site, and the $C^{\delta+}$ atom toward the middle point of two adjacent $Cl^-$ sites. The "F" minimum on the 2×2-MBD PES is -4094.4 $cm^{-1}$. The landscape for the 1×1-MBD and 1×1-D3 PESs is similar, as shown in Figures 7(c) and 7(d), except the barrier between the wells is higher. The PBE-MBD@rsSCS/FI and PBE-D3(BJ) functional give similar results in both $p(1×1)$ and $p(2×2)$ models, except the PBE-D3 interaction energies are larger.

It is noted that the interaction for vibrationally excited CO with the NaCl surface is much stronger than that for vibrationless CO. More importantly, the C-down configuration prevailed near the CO equilibrium geometry is no longer the most stable. This is likely due to the peculiar feature of CO ES properties. For the CO molecule at its equilibrium geometry, the dipole moment is 0.12 Debye in a direction of $C^{\delta-}O^{\delta+}$. When $r_{CO}$ is increased to 1.596 Å, the dipole moment changes its direction to $C^{\delta+}O^{\delta-}$, and has a much larger value of 1.14 Debye (calculated at AE-CCSD(T)/aug-cc-pCV5Z level).[45] This explains why the interaction of a stretched CO molecule with the NaCl(100) surface is much stronger than that of equilibrium CO. This change of the CO dipole with the CO bond length was identified as the driving force for the flipping of vibrationally excited CO on NaCl(100) observed in the recent experiment.[21] However, we emphasize that the parallel geometries shown in Figure 7 are not particularly relevant to the experiment because the periodicity of these models dictates that all CO adsorbates have the same orientation and bond-length. In experiment, the vibrationally excited CO is likely surrounded by adsorbates with no or low vibrational excitations in C-down configurations. This important difference will be discussed in more detail below.

**C. Isomerization Pathway**

The minimum energy paths (MEPs) connecting from the C-down and O-down minima has been determined both on the 2×2-MBD PES and 1×1-MBD PES, as plotted in Figure 8. Snapshots along the MEPs can be visualized in movies in SI. Both the MEPs on the 2×2-MBD and 1×1-MBD PES feature the flipping of CO accompanying a significant CO diffusion from one $Na^+$ site to an adjacent $Na^+$ site. On the 1×1-MBD PES, the transition state locates at a $Cl^-$ site with the tilt angle of 88°, nearly parallel to the surface. The azimuthal angle keeps at -135° along the diffusion/isomerization path. The C-down *s*2/*s*3 and the O-down *s*4 are clearly shown in the MEP. The barrier has an energy of -653.4 $cm^{-1}$ relative to the dissociation limit, which is 848.4 $cm^{-1}$ higher than the global C-down minimum. This indicates that the isomerization needs not proceed via desorption. However, the MEP may not be relevant to the recent experiments[20, 21] because all CO adsorbates on the surface are assumed to diffuse in sync, due to the periodic conditions, which is unlikely to happen at the coverage of 1 ML. On the 2×2-MBD PES, the transition state locates at the middle point of two adjacent $Cl^-$ sites, with a tilt angle of 110° and azimuthal angle of 180°. The energy is -679.4 and 701.9 $cm^{-1}$, relative to the dissociation limit and the global C-down minimum, respectively. This scenario is also not relevant to the recent experiments[20, 21] because of its lower coverage.

We consider the collective diffusion associated with isomerization as an artifact of the models, as alluded above. To gain a better understanding of the flipping without the diffusion, we examine the isomerization paths on the two PESs with the center-of-mass of CO fixed at top of the $Na^+$ site, which are displayed as dashed lines in Figure 8. The isomerization barrier height is 943.2 cm$^{-1}$ on the 1×1-MBD PES, and 1080.6 cm$^{-1}$ on the 2×2-MBD PES, which are 94.8 and 378.7 cm$^{-1}$ higher than the diffusion/isomerization paths, for the two PESs, respectively.

It is seen from Tables 1 and 3 that the vibrational frequency of CO shows a small blue shift at the C-down minima, and a small red shift at the O-down minima, in both the $p(1\times1)$ (1 ML) and $p(2\times2)$ (1/4 ML) models. These frequency shifts are responsible for the switch of energy order of the two configurations at highly excited CO vibrational states.[21] To illustrate this point, the vibrationally adiabatic potential energy curves ($V_a^v$) along the isomerization reaction paths for different CO vibrational quantum numbers ($v$) up to 40 are calculated, assuming the center-of-mass of CO is fixed at the Na$^+$ site. These curves are generated by computing the vibrational bound states at each tilt angle on a one-dimensional PEC of CO, which is obtained from varying $r_{CO}$ of the corresponding point along the MEP and optimizing $dZ$ on the full-dimensional PES, with tilt and azimuthal angles fixed at the original value and center-of-mass of CO fixed at the Na$^+$ site. From Figure 9, one can see that, with the increasing of vibrational quantum number, the energy difference between O-down and C-down structures decreases. At $v = 30$, the O-down structure becomes more energetically favorable than its C-down counterpart. This provides the energetic driving force for the flipping of vibrationally excited CO adsorbate observed in the experiment.[21] The barrier height separating the C-down and O-down minima also decreases as the vibrational quantum number increases, consistent with the PES landscape of stretched CO (Figure 7). Results on the 2×2-MBD PES and the 1×1-MBD PES are qualitatively similar to each other. This observation is consistent with experimentally observed trends.[20, 21] Quantitatively, the CO frequency shifts for the C-down and O-down configurations are +4.9 and -7.5 cm$^{-1}$, using the $p(1\times1)$ model with PBE-MBD@rsSCS/FI functional for $^{13}$C$^{18}$O, which can compared with the experimental values of +7.6 and -9.3 cm$^{-1}$. A complete list of frequency shifts at various C-down and O-down configurations is given in Table 4.

It is interesting to note that the *v*-dependent isomerization potentials shown in Figure 9 are qualitatively consistent with the results of a simple model that only considers the ES interactions of CO fixed at a position above the surface with the electric field generated by the NaCl substrate.[21] The predominate ES interaction leading to the change of the energy order of the C-down and O-down configurations was attributed to the switch of the CO dipole as a function of the vibrational excitation. This simple model underscores the dominant nature of the CO-NaCl ES interaction, but unfortunately it cannot be used to describe the actual dynamics of isomerization due to the absence of vdW interactions and short-range repulsion. The DFT PESs developed in this work contain ES, vdW and short-range interactions, and thus amenable to characterization of not only adsorption, but also isomerization dynamics.

**IV. Conclusions**

The PESs constructed from vdW-corrected functionals provide valuable insights into the adsorption and isomerization of CO molecules on NaCl(100). The general features of the experimentally observed *p*(2×1) CO adsorption pattern and the adsorption energy are reasonably reproduced. The PESs reveal a rather flat energy landscape near the equilibrium adsorption geometry, stemming apparently from the weak adsorbate-adsorbate and adsorbate-substrate interactions. Concerning the isomerization of vibrationally excited CO, the PESs clearly show the C-down and O-down potential minima and the isomerization pathway between them. They also qualitatively reproduce the frequency shifts in the C-down and O-down configurations and confirm them as the origin of the flipping of the vibrationally excited CO adsorbate.

Despite the insights they provide, however, these PESs reported in this work cannot be directly used to model the experiment conducted for a monolayer of CO in the p(2x1) structure before vibrational excitation.[20, 21] This is due to the enforced periodicity in these PESs, which requires the adjacent adsorbates to have the same position, orientation, and vibrational excitation as the ones in the unit cell. This is obviously not the situation in the experiment,[20, 21] where a highly excited CO is most likely surrounded by co-adsorbates with potentially different coordinates and orientations, as well as no or low vibrational excitations. However, the PESs, represent the first step towards a realistic simulation of the experiment. To that end, we envisage a sufficiently large unit cell in which one or few vibrationally excited CO adsorbates are surrounded by multiple CO adsorbates that are mostly in their low-lying vibrational states. Thanks to the weak interacting nature of the system, the total interaction energy can in principle be decomposed into pairwise molecule-molecule and molecule-surface interactions, with negligible many-body terms. The former can be constructed with two isolated CO molecules, as we did recently,[45] accounting for both the short-range vdW and long-range electrostatic interactions. The latter can be obtained using the $p(2\times2)$ model or a model with even larger surface unit cell, in which the interaction amongst the CO adsorbates is practically zero. Work in this direction is already underway. Once such a composite PES become available, we can start to simulate the energy transfer as well as the isomerization dynamics.

**Acknowledgements**: This work is supported by a grant from the Air Force Office of Scientific Research (Grant No. FA9550-15-1-0305). J. C. thanks partial support from the National Natural Science Foundation of China (Grant No. 21803046). S. H. and J. M. acknowledge financial support from the Leiden Institute of Chemistry (LIC) and the Netherlands Organisation for Scientific

Research (NWO) under Vidi Grant No. 723.014.009. H. G. is a Humboldt Research Awardee and thanks Prof. Alec Wodtke for his warm hospitality during visits to Göttingen and Profs. Alec Wodtke and Dirk Schwarzer and Dr. Jascha Lau for several stimulating discussions. He also thanks Prof. Joel Bowman for several useful discussions. The calculations were performed at the Center for Advanced Research Computing (CARC) at UNM.

Table 1. Adsorption energies, geometries, and CO frequencies for minima optimized in the $p(1\times1)$ surface unit cell (1 ML) using different functionals. The global minimum in each case is given in bold. The harmonic frequencies for free CO are 2131.8, 2134.3 and 2130.5 cm$^{-1}$ for the three functionals, respectively.

| Method | Geometry | Adsorption energy (cm$^{-1}$) | $v_{CO}$ (cm$^{-1}$) | $r_{CO}$ (Å) | $dZ^*$ (Å) | Tilt (°) | Azimuthal (°) | $dY^{**}$ (Å) | $dX^{**}$ (Å) |
|---|---|---|---|---|---|---|---|---|---|
| PBE | C-down s1 | **-885.39** | 2132.58 | 1.1352 | 3.1581 | 30.19 | 1.96 | 0.0447 | 0.6588 |
|  | C-down s2 | -883.77 | 2138.32 | 1.1355 | 3.1946 | 27.37 | 45.00 | 0.4270 | 0.4270 |
|  | C-down s3 | -818.79 | 2136.91 | 1.1349 | 3.2826 | 2.93 | 44.56 | 0.0405 | 0.0399 |
|  | O-down s4 | -352.24 | 2125.32 | 1.1360 | 3.3512 | 143.39 | -135.04 | 0.1930 | 0.2114 |
|  | O-down s5 | -312.04 | 2135.64 | 1.1358 | 3.3468 | 180 | - | 0 | 0 |
| PBE-D3(BJ) | C-down s1 | **-1996.07** | 2139.18 | 1.1353 | 3.0412 | 36.13 | 3.77 | 0.0557 | 0.8328 |
|  | C-down s2 | -1974.58 | 2137.44 | 1.1351 | 3.1061 | 32.16 | 45.00 | 0.5200 | 0.5200 |
|  | C-down s3 | -1977.76 | 2132.72 | 1.1352 | 3.1068 | 33.72 | 39.34 | 0.5033 | 0.5886 |
|  | O-down s4 | -1433.25 | 2125.80 | 1.1365 | 3.0558 | 122.78 | -135.15 | 0.2696 | 0.2656 |
| PBE-MBD@rsSCS/FI | C-down s1 | **-1501.79** | 2135.55 | 1.1348 | 3.1128 | 31.00 | 3.28 | 0.0495 | 0.6472 |
|  | C-down s2 | -1490.89 | 2134.05 | 1.1348 | 3.1439 | 27.95 | 45.00 | 0.4189 | 0.4189 |
|  | C-down s3 | -1416.46 | 2150.18 | 1.1345 | 3.2505 | 7.61 | 45.00 | 0.0617 | 0.0614 |
|  | O-down s4 | -909.44 | 2122.67 | 1.1360 | 3.1982 | 141.37 | -136.60 | 0.1779 | 0.2026 |

$^*$ The height of CO to surface $dZ$ is defined as $(dZ_C+dZ_O)/2$.
$^{**}$ The lateral displacements $dX/dY$ correspond to $dX_C/dY_C$ for C-down configurations, and $dX_O/dY_O$ for O-down configurations.

Table 2. Adsorption energies, geometries, and CO frequencies for minima optimized in the $p(2\times1)$ surface unit cell (1 ML) using different functionals. The global minimum in each case is given in bold. The harmonic frequencies for free CO are 2131.8, 2134.3 and 2130.5 cm$^{-1}$ for the three functionals, respectively. The experimental structure information[12] is included for comparison.

| Method | Geometry | Adsoprtion energy (cm$^{-1}$) | $\nu_{CO}$ (cm$^{-1}$) | $r_{CO}$(Å) | $dZ^*$ (Å) | Tilt (°) | Azimuthal (°) | $dY^{**}$ (Å) | $dX^{**}$ (Å) |
|---|---|---|---|---|---|---|---|---|---|
| PBE | C-down s1 | -891.30 | 2131.8/2124.2 | 1.1350 | 3.1391 | 31.83 | 86.56 | 0.7163 | 0.0487 |
| | C-down s2 | -884.92 | 2135.0/2128.8 | 1.1355 | 3.2133 | 25.67 | 45.92 | 0.4044 | 0.3757 |
| | C-down s3 | **-902.70** | 2131.1/2123.9 | 1.1351 | 3.1375 | 33.48 | ±89.42 | ±0.7633 | 0.0152 |
| | C-down s4 | -889.98 | 2134.1/2128.0 | 1.1354 | 3.2085 | 25.66 | ±47.30 | ±0.4116 | 0.3765 |
| | C-down s5 | -892.58 | 2136.3/2131.7 | 1.1352 | 3.1918 | 27.27 | ±21.21 | ±0.2217 | 0.5647 |
| | O-down s6 | -352.99 | 2127.0/2126.7 | 1.1357 | 3.3078 | 142.13 | -146.76 | 0.1252 | 0.3191 |
| | O-down s7 | -352.67 | 2123.6/2123.1 | 1.1357 | 3.3222 | 144.01 | ±153.07 | ±0.1348 | 0.3031 |
| PBE-D3(BJ) | C-down s1 | -2004.20 | 2138.2/2130.8 | 1.1353 | 3.0454 | 35.83 | 87.11 | 0.8152 | 0.0327 |
| | C-down s2 | -1988.55 | 2139.6/2134.6 | 1.1363 | 3.0418 | 35.83 | 48.65 | 0.6134 | 0.5631 |
| | C-down s3 | **-2008.54** | 2134.3/2126.8 | 1.1354 | 3.0462 | 35.42 | ±87.08 | ±0.8032 | 0.0431 |
| | C-down s4 | -1987.13 | 2134.5/2129.2 | 1.1342 | 3.1060 | 31.38 | ±44.40 | ±0.5177 | 0.5121 |
| | C-down s5 | -1999.73 | 2133.9/2131.3 | 1.1345 | 3.0905 | 31.47 | ±20.83 | ±0.2281 | 0.6940 |
| | O-down s6 | -1406.80 | 2130.5/2129.5 | 1.1358 | 3.1583 | 137.93 | -146.76 | 0.0933 | 0.2840 |
| | O-down s7 | -1403.73 | 2127.2/2125.2 | 1.1366 | 3.1415 | 134.51 | ±153.71 | ±0.0885 | 0.3972 |
| PBE-MBD@rsSCS/FI | C-down s1 | -1542.17 | 2138.6/2130.9 | 1.1349 | 3.0590 | 33.83 | 87.03 | 0.7665 | 0.0215 |
| | C-down s2 | -1531.22 | 2134.8/2128.7 | 1.1354 | 3.1240 | 28.33 | 46.02 | 0.4451 | 0.4363 |
| | C-down s3 | **-1549.95** | 2133.8/2126.3 | 1.1353 | 3.0558 | 34.43 | ±87.51 | ±0.7737 | 0.0279 |
| | C-down s4 | -1523.92 | 2137.4/2131.1 | 1.1340 | 3.1467 | 26.16 | ±46.94 | ±0.3932 | 0.3781 |
| | C-down s5 | -1534.89 | 2137.3/2133.2 | 1.1353 | 3.1112 | 29.65 | ±17.40 | ±0.2022 | 0.6070 |
| | O-down s6 | -942.587 | 2129.9/2129.2 | 1.1359 | 3.1459 | 140.58 | -148.24 | 0.1204 | 0.2967 |
| | O-down s7 | -943.903 | 2130.2/2129.6 | 1.1360 | 3.1463 | 140.50 | ±149.19 | ±0.1292 | 0.2895 |
| Expt. | C-down | - | - | 1.1±0.1 | 3.08±0.15 | 28±5 | 30±50 | 0.4±0.2 | 0.1±0.3 |

$^*$ The height of CO to surface $dZ$ is defined as $(dZ_C+dZ_O)/2$.
$^{**}$ The lateral displacements $dX/dY$ correspond to $dX_C/dY_C$ for C-down configurations, and $dX_O/dY_O$ for O-down configurations.

Table 3. Adsorption energies, geometries, and CO frequencies for minima optimized in the $p(2\times2)$ surface unit cell (1/4 ML) using different functionals. The global minimum in each case is given in bold. The harmonic frequencies for free CO are 2131.8, 2134.3 and 2130.5 cm$^{-1}$ for the three functionals, respectively.

| Method | Geometry | Adsorption energy (cm$^{-1}$) | $v_{CO}$ (cm$^{-1}$) | $r_{CO}$ (Å) | $dZ^*$ (Å) | Tilt (°) | Azimuthal (°) | $dY^{**}$ (Å) | $dX^{**}$ (Å) |
|---|---|---|---|---|---|---|---|---|---|
| PBE | C-down s1 | -887.08 | 2131.46 | 1.1349 | 3.1353 | 30.34 | 2.12 | 0.0341 | 0.7164 |
|  | C-down s2 | -900.15 | 2136.19 | 1.1346 | 3.1822 | 23.78 | 45.00 | 0.3983 | 0.3983 |
|  | **C-down s3** | **-912.44** | 2138.73 | 1.1342 | 3.3207 | 0.60 | 0 | 0.0010 | 0 |
|  | O-down s4 | -351.43 | 2124.66 | 1.1357 | 3.2118 | 169.74 | -135.00 | 0.0822 | 0.0822 |
|  | O-down s5 | -367.30 | 2131.33 | 1.1360 | 3.3351 | 180 | - | 0 | 0 |
| PBE-D3(BJ) | **C-down s1** | **-1767.55** | 2140.76 | 1.1344 | 3.0671 | 31.16 | 1.68 | 0.0415 | 0.7267 |
|  | C-down s2 | -1748.42 | 2141.84 | 1.1344 | 3.0846 | 27.95 | 45.00 | 0.4963 | 0.4963 |
|  | C-down s3 | -1715.79 | 2145.05 | 1.1338 | 3.2459 | 0.01 | 0 | 0.0001 | 0 |
|  | O-down s4 | -1102.07 | 2128.97 | 1.1354 | 3.1813 | 169.72 | -135.00 | 0.0846 | 0.0846 |
| PBE-MBD@rsSCS/FI | **C-down s1** | **-1381.28** | 2134.27 | 1.1341 | 3.2019 | 17.66 | 0 | 0 | 0.3900 |
|  | C-down s2 | -1376.63 | 2138.23 | 1.1343 | 3.1559 | 23.56 | -135.00 | 0.3950 | 0.3950 |
|  | C-down s3 | - | - | - | - | - | - | - | - |
|  | O-down s4 | - | - | - | - | - | - | - | - |
|  | O-down s5 | -784.67 | 2123.37 | 1.1357 | 3.2448 | 180 | - | 0 | 0 |
|  | C-down s6 | -1379.56 | 2147.67 | 1.1339 | 3.2466 | 0 | - | 0 | 0 |

$^*$ The height of CO to surface $dZ$ is defined as $(dZ_C+dZ_O)/2$.
$^{**}$ The lateral displacements $dX/dY$ correspond to $dX_C/dY_C$ for C-down configurations, and $dX_O/dY_O$ for O-down configurations.

Table 4. Frequency shifts relative to the free $^{13}C^{18}O$ for minima optimized in the $p(1\times1)$ and $p(2\times2)$ surface unit cells using different functionals. The global minimum and energetically most favorable O-down minimum in each case are given in bold.

|  | $p(1\times1)$ (1 ML) | | | $p(2\times2)$ (1/4 ML) | | |
|---|---|---|---|---|---|---|
|  | PBE | PBE-D3(BJ) | PBE-MBD@rsSCS/FI | PBE | PBE-D3(BJ) | PBE-MBD@rsSCS/FI |
| C-down *s1* | **0.73** | **4.89** | **4.87** | -0.34 | **6.41** | **3.53** |
| C-down *s2* | 6.31 | 3.05 | 3.40 | 4.16 | 7.39 | 7.50 |
| C-down *s3* | 4.80 | -1.46 | 18.95 | **6.60** | 10.46 |  |
| O-down *s4* | **-6.19** | **-7.81** | **-7.51** | -6.92 | **-5.14** |  |
| O-down *s5* | 3.80 |  |  | **-0.31** |  | **-6.99** |
| C-down *s6* |  |  |  |  |  | 16.65 |

Figure 1. Spatial and energy distributions of all the configurations used for PES construction and the definition of the coordinates used to describe the geometries. (a) Distribution on the $p(1\times1)$ unit cell; (b) Distribution on the $p(2\times2)$ unit cell; (c) Spatial distribution on the height of the CO molecule to the surface and $r_{CO}$; (d) Potential energy ($2\times2$-MBD) as a function of $r_{CO}$; (e)-(f) Definition of the coordinates in top (e) and side (f) views. The $Na^+$ and $Cl^-$ are represented by blue and yellow circles, respectively.

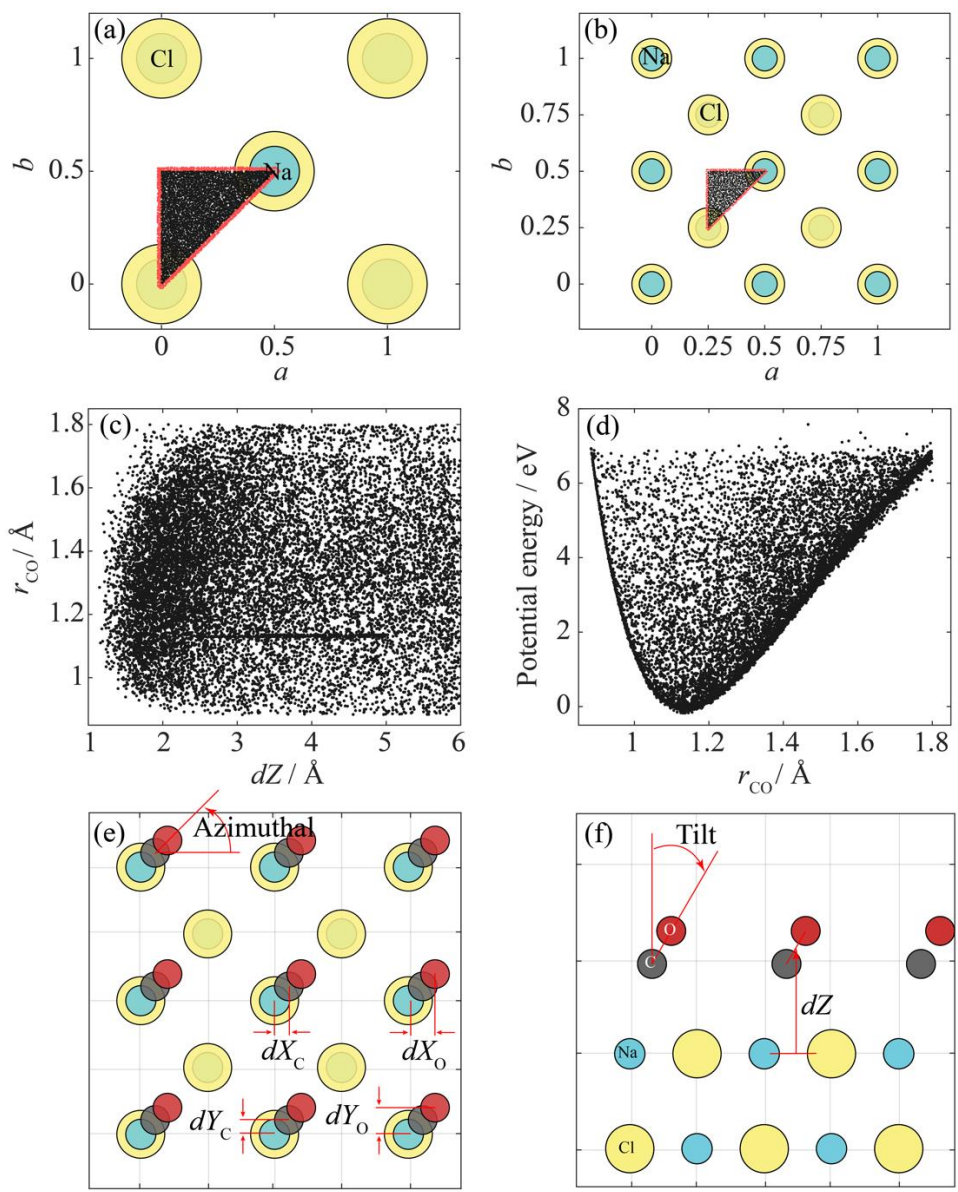

Figure 2. Illustration of geometries C-down *s*1/2/3 and O-down *s*4 optimized by the PBE functional in the *p*(1×1) and *p*(2×2) unit cells. The Na$^+$, Cl$^-$, O and C are represented by blue, yellow, red and gray circles, respectively.

*p*(1×1) C-down *s*1

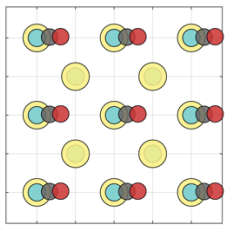 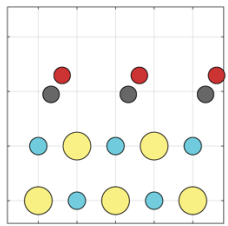

*p*(2×2) C-down *s*1

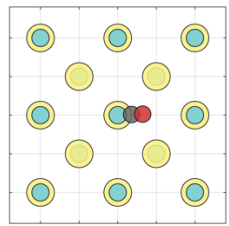 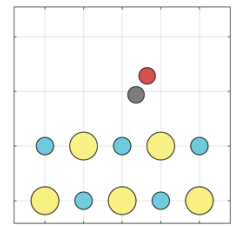

*p*(1×1) C-down *s*2

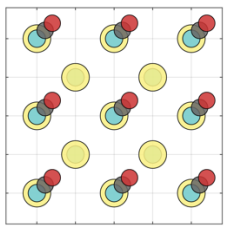 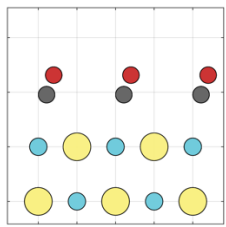

*p*(2×2) C-down *s*2

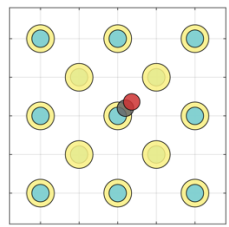 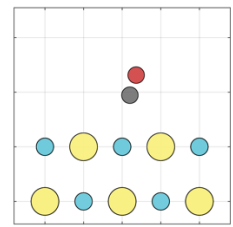

*p*(1×1) C-down *s*3

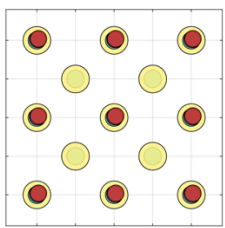 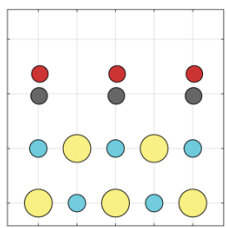

*p*(2×2) C-down *s*3

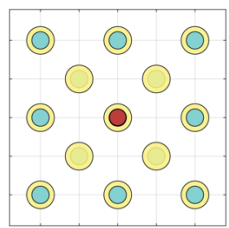 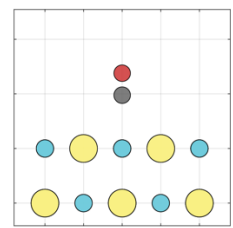

*p*(1×1) O-down *s*4

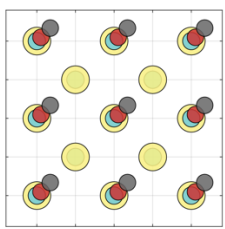 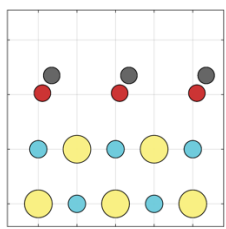

*p*(2×2) O-down *s*4

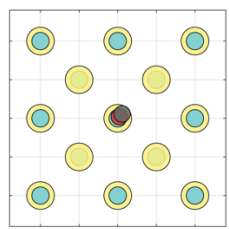 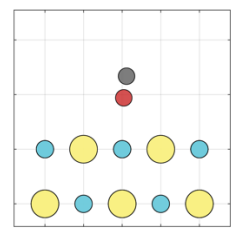

Figure 3. Illustration of geometries C-down *s*1-*s*5 and O-down *s*6/7 in the *p*(2×1) unit cell. Results from different functionals are similar. The Na$^+$, Cl$^-$, O and C are represented by blue, yellow, red and gray circles, respectively.

*p*(2×1) C-down *s*1 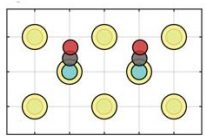 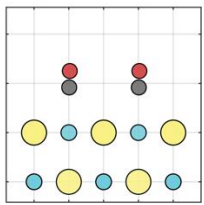  *p*(2×1) C-down *s*2 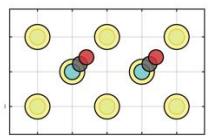 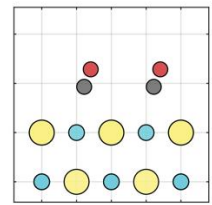

*p*(2×1) C-down *s*3 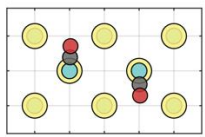 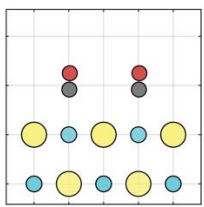  *p*(2×1) C-down *s*4 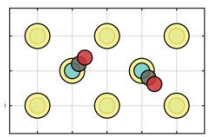 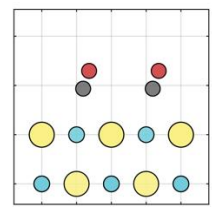

*p*(2×1) C-down *s*5 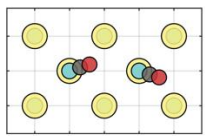 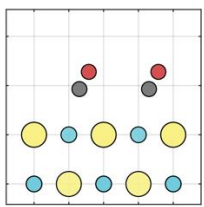  *p*(2×1) O-down *s*6 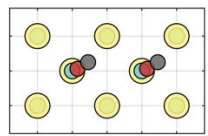 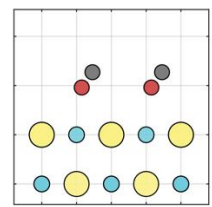

*p*(2×1) O-down *s*7 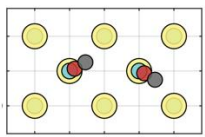 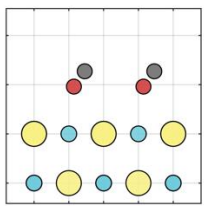

Figure 4. Adsorption potential energy curves from different methods as well as the comparison with experimental values.

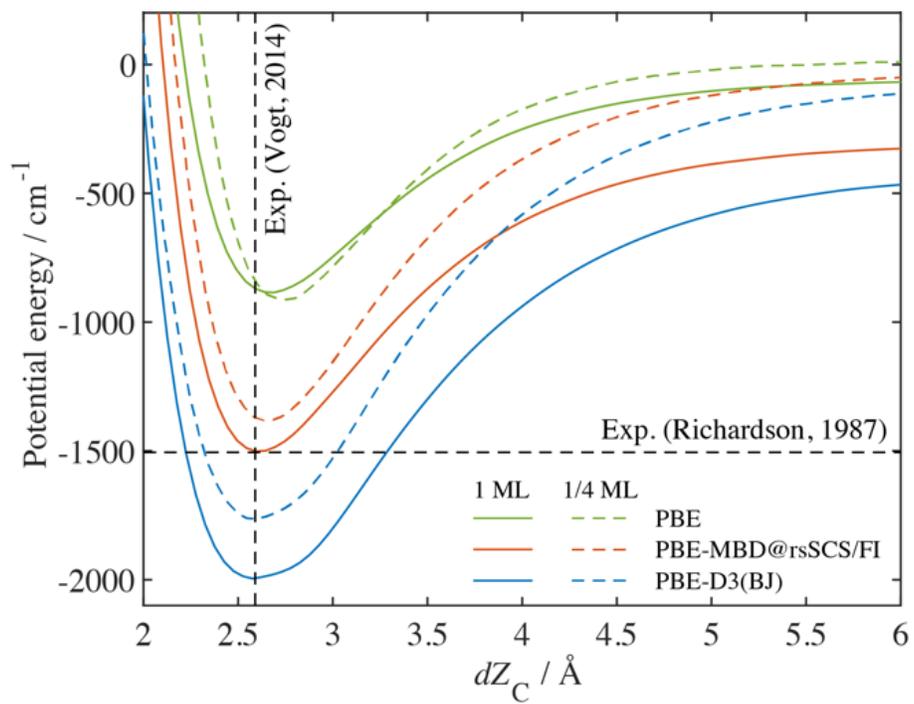

Figure 5. Contours of PES as a function of the tilt angle and the height of the CO molecule to the surface, with $r_{CO}$ fixed at equilibrium (1.132 Å), and the other three coordinates optimized. (a) to (d) correspond to the 2×2-MBD, 2×2-D3, 1×1-MBD and 1×1-D3 PESs, respectively. The red markers "A" to "D" are the global C-down and O-down minima.

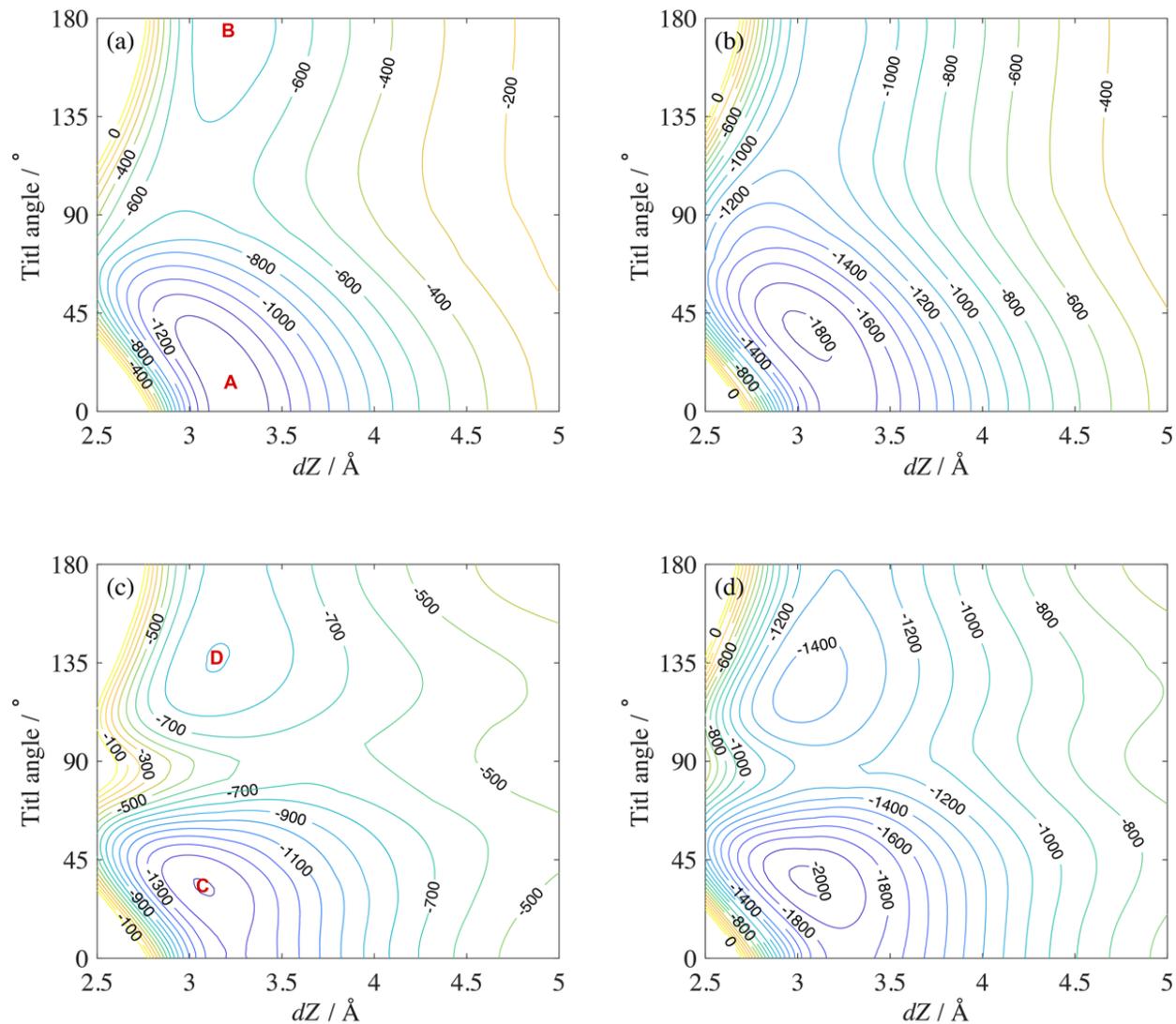

Figure 6. Contours of the PES for CO diffusion among different sites, with $r_{CO}$ fixed at the equilibrium (1.132 Å), and the other three coordinates optimized. (a) the 2×2-MBD PES; and (b) the 1×1-MBD PES. The red markers "$s1$" to "$s3$" correspond to the tilted C-down minima defined in Tables 1 and 3.

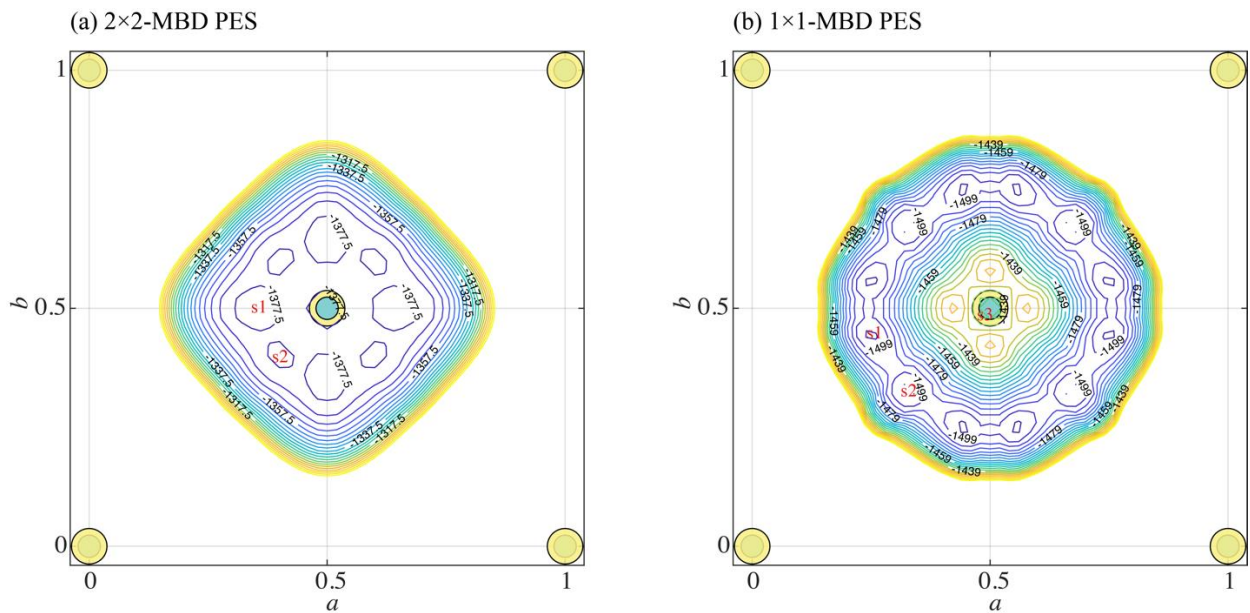

Figure 7. Same as Fig. 4, but with $r_{CO}$ fixed at 1.596 Å which is the outer turning point of CO($v$=20). The red markers "E" to "H" correspond to the global and local "minima".

Figure 8. Minimum energy paths connecting the C-down and O-down minima on the 1×1-MBD and 2×2-MBD PESs. The dashed lines correspond to the center-of-mass of CO fixed at the Na$^+$ site.

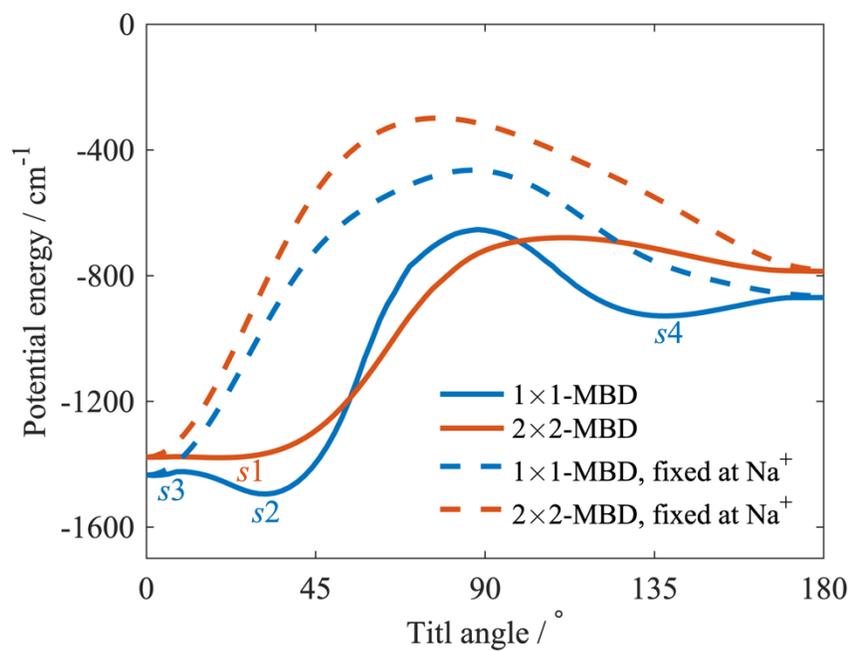

Figure 9. Vibrational adiabatic potential energy curves along the isomerization MEPs with the center-of-mass of CO fixed at the Na$^+$ site. (a) The 2×2-MBD PES; (b) The 1×1-MBD PES.

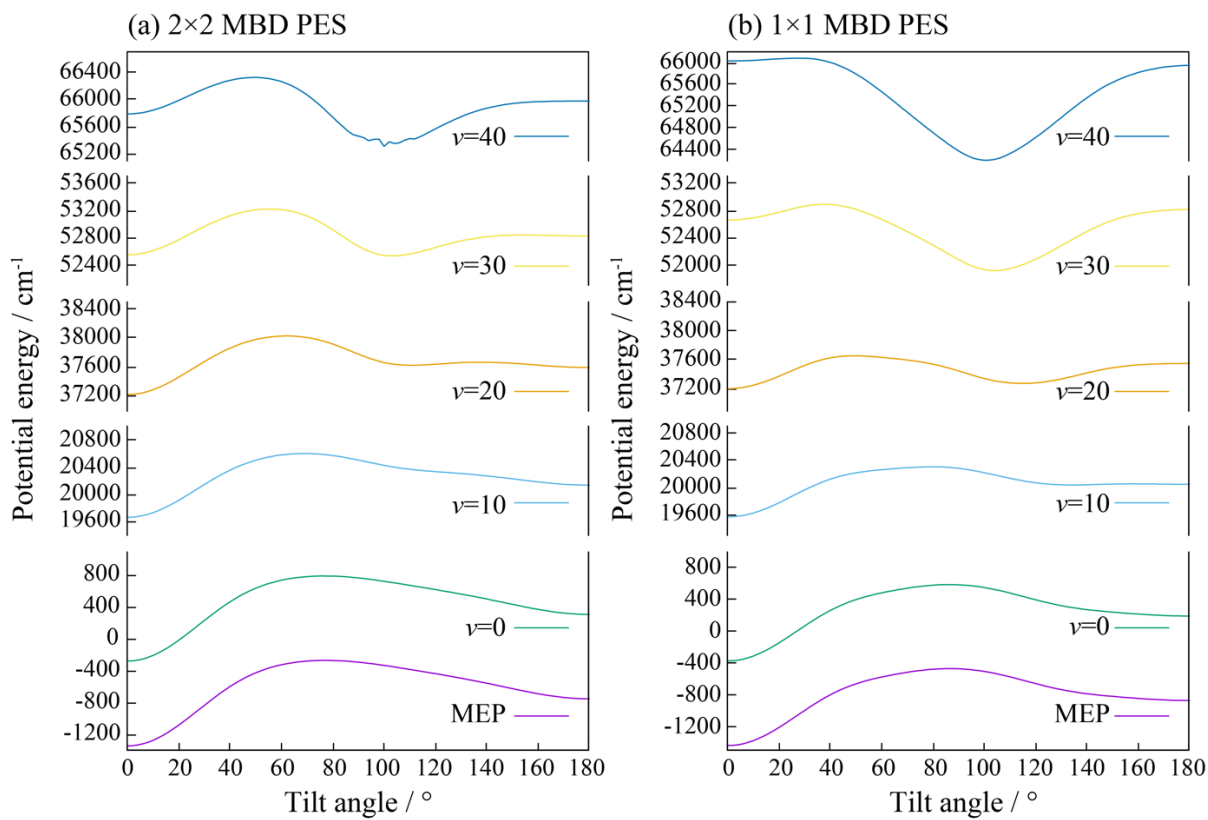